\documentclass{eas}
\usepackage{graphicx}

\begin{document}

\title{Hi-fi phenomenological description of eclipsing binary light variations as the basis for their period analysis}
\author{Zden\v{e}k Mikul\'{a}\v{s}ek}
\address{Department of Theoretical Physics and Astrophysics, Masaryk University, Brno, Czech Republic; \email{mikulas@physics.muni.cz}}
\secondaddress{Observatory \& Planetarium of J.~Palisa, V\v SB--Technical University, Ostrava, Czech Republic}
\author{Miloslav Zejda}\sameaddress{1}
\author{Marek Chrastina}\sameaddress{1}
\author{ShengBang~Qian}\address{National Astronomical Observatories/Yunnan Observatory of Chinese Academy of Sciences, Kunming, China}
\secondaddress{Key Laboratory for the Structure and Evolution of Celestial Objects, Chinese Academy of Sciences, Kunming, China}
\author{LiYing Zhu}
\sameaddress{3,\,4}
\begin{abstract}
In-depth analysis of eclipsing binary (EB) observational data collected for several decades can inform us about a lot of astrophysically interesting processes taking place in the systems. We have developed a wide-ranging method for the phenomenological modelling of eclipsing binary phase curves that enables us to combine even very disparate sources of phase information. This approach is appropriate for the processing of both standard photometric series of eclipses and data from photometric surveys of all kind.

We conclude that mid-eclipse times, determined using the latest version of our `hi-fi' phenomenological light curve models, as well as their accuracy, are nearly the same as the values obtained using much more complex standard physical EB models.
\end{abstract}
\runningtitle{Z. Mikul\'a\v{s}ek \etal: Hi-fi description of eclipsing binary light variations}
\maketitle
\section{Introduction}\label{intr}

\begin{figure}
\centering \resizebox{0.50\hsize}{!}{\includegraphics{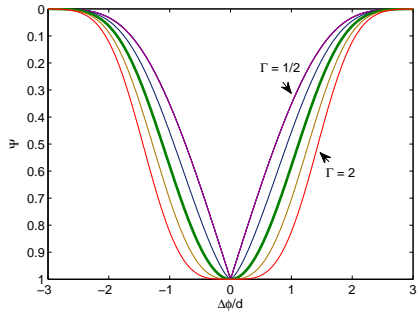}}
\caption{\small{Examples of light curve during eclipses allowed by our phenomenological EB models. $d$ is a parameter describing the duration of the eclipse, $\Gamma$ is a parameter expressing the kurtosis of the light curve.}} \label{cecka}
\end{figure}

There are two groups of astrophysical tasks standardly solved by the analysis of eclipsing binary (EB) observations. We can derive the physics of the present state of a system and its components -– namely dimensions and geometry of the system, outer characteristics of both stars such as radii, masses, temperatures, limb darkening, and parameters of possible streams and disks. Information is extracted by means of sophisticated physical models of the double star applied to numerous and precise data of all kind obtained by contemporary observational methods and approaches developed for this purpose. Equally important is the evolution of the system in the time scale of decades which may manifest itself in the gradual changes of some characteristics of the system, namely its observed period of phase variations. These usually very tiny changes may inform us about e.g. the rate of mass exchange between the interacting components, the presence of a possible invisible star or planet in the system, we can test the interior structure of the components, etc. Such period analyses can be also treated by the simple, physically learned phenomenological models, applied to both present and archival data of sometimes very suspicious quality and properties.

The two approaches to eclipsing binary research differ not only with respect to their data processing and modelling, but also with respect to their data requirements.While the first approach needs the most accurate and homogeneous data possible and obtained in a short time interval where the periods can be put constant (ideal data are from extraterrestrial observatories like Kepler, MOST and other), the second approach prefers observations that need not be of superior quality, but that should be spread over the longest possible interval.

\section{Phenomenological models of eclipsing binary light curves} \label{phenomod}

During the last few decades we have concentrated on the modelling of complete light curves of rotating (e.g. magnetic chemically peculiar stars) and orbiting (eclipsing binaries or stars with transiting planets) variables by periodic functions that can be sufficiently well described by a minimum of free parameters. To select the model functions for eclipsing binaries, we consider that they have to describe as aptly as possible those parts of light curves (LC) that are in the vicinity of their inflex points, where their slopes are maximal -- hence LCs during eclipses and transits.

\begin{figure}
\centering \resizebox{0.52\hsize}{!}{\includegraphics{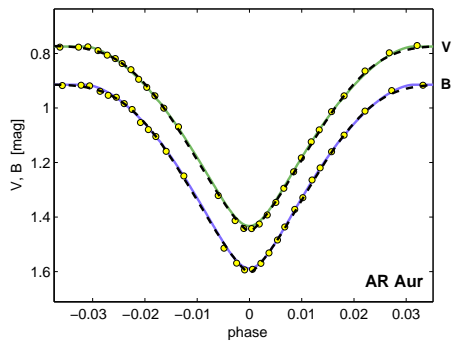}}
\caption{\small{The fit of the primary minimum of the detached eclipsing binary AR Aur in $V$ and $B$ filters, represented by their normal points (yellowy filled circles), by physical (full lines) and phenomenological (dashed lines) light curves.}} \label{arprim}
\end{figure}

We have established a three-parametric symmetrical function parameterized by its amplitude, half-width and the parameter of the LC kurtosis $\Gamma$ that enables the approximation of a wide variety of real EB LCs (see also Fig.\,\ref{cecka}). Such model function is suitable for the fitting of light curves during transits as well as that of asymmetric light curves that show proximity effects (asphericity of components, reflection) or the standard O'Connell effect. We can also with high-fidelity simply simulate LCs of eclipsing binaries moving in orbits with an eccentricity of less than 0.25. The modular architecture of our phenomenological LC models has the advantage that all observations containing phase information, including photometry done in various filters, non-complete observational sets and surveys of disparate quality, can be processed at once. In addition it allows them to be combined with e.g. radial velocity measurements.

\section{Comparison of results of standard and phenomenological methods} \label{compar}

The indisputable advantage of the presented phenomenological method (PM) is its clarity and intuitiveness in use. Nevertheless, it should also give reliable results burdened by uncertainties comparable with those obtained by standard methods. The basic data used in period analyses of EBs are the \emph{O-C} values and their uncertainties obtained from observational sets.

In the scope of our pilot project we selected several eclipsing binaries of diverse LC types listed in the CALEB, (see http://caleb.eastern.edu/). We used individual $V$ measurements from this source and evaluated the times of the zero phases derived from the fit of our model with the minimum of free parameters and the same quantities derived from the fit of the Binary Maker\,3 synthetic curve. We found that the results themselves are almost the same, as well as their uncertainties. The detailed description and the discussion of the system of our phenomenological eclipsing binary modelling will be published elsewhere in the near future.

\begin{figure}
\centering \resizebox{0.52\hsize}{!}{\includegraphics{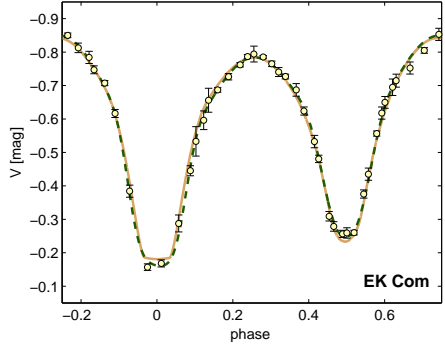}}
\caption{\small{The fit of the $V$ light curve of the overcontact eclipsing binary EK Com. The $V$ observations are represented by normal points with error bars. The synthetic model curve is depicted by a solid line, the phenomenological LC is signed by the dash line.}} \label{EKCom}
\end{figure}

\subsection{Eclipsing binaries AR Aurigae and EK Comae} \label{AREK}
The findings can be illustrated using two well-known different EBs.

\textbf{AR Aurigae} is a detached eclipsing binary, the period $P=4.13169$\,d, $r_1=0.0977,\ r_2=0.0996,\ T_{\mathrm{eff}1}=11\,100$\,K, $T_{\mathrm{eff}2}=10\,600$\,K, $i=88.5^{\circ},\ \varepsilon= 0.00$, $M_2/M_1=0.922$ (\cite{nord}), $B$ and $V$ photometry of O'Connell \cite{con}. For the description of both light curves a model with only \textbf{9}\,free parameters was used (see Fig.\,\ref{arprim}), $\delta_{\mathrm{phen}}=6.82\times10^{-5}$\,d, $\delta_{\mathrm{phys}}=6.81\times10^{-5}$\,d.

\textbf{EK Comae} is an overcontact, spotted eclipsing binary with an ultra-short orbital period of $P=0.26669$\,d, $r_1=0.524,\ r_2=0.318,\ T_{\mathrm{eff}1}=5\,000$\,K, $T_{\mathrm{eff}2}=5\,300$\,K, $i=88.5^{\circ},\ \varepsilon= 0.00$, $M_2/M_1=0.922$ (\cite{samec}). For the description of the $V$ LC the simple phenomenological model with only \textbf{8}\,free parameters was used (see Fig.\,\ref{EKCom}), $\delta_{\mathrm{phen}}=1.1\times10^{-4}$\,d,  $\delta_{\mathrm{phys}}=1.3\times10^{-4}$\,d.

\section{Conclusions} \label{concl}

We conclude that our phenomenological models of eclipsing binary light curves are able to replace those derived from physical models in common period analyses of photometry obtained at ground-based observatories.
\\
\\
\noindent \emph{Acknowledgements:} The investigation was supported
by projects LH12175 and GA \v{C}R 13-10589S. Authors thank S. de Villiers for the language revision.

\end{document}